\documentclass[showpacs,aps,prl,twocolumn,superscriptaddress]{revtex4}
\usepackage{graphicx} %graphics handler
\usepackage{dcolumn}% Align table columns on decimal point
\usepackage{bm}% bold math
\usepackage{amssymb,amsmath}
\usepackage{epstopdf}
\usepackage[T1]{fontenc}
\usepackage[latin9]{inputenc}
\usepackage{color}
\usepackage{float}
\usepackage{esint}
\usepackage{graphics}
\usepackage{bm}
\makeatletter
\@ifundefined{textcolor}{}
{%
 \definecolor{BLACK}{gray}{0}
 \definecolor{WHITE}{gray}{1}
 \definecolor{RED}{rgb}{1,0,0}
 \definecolor{GREEN}{rgb}{0,1,0}
 \definecolor{BLUE}{rgb}{0,0,1}
 \definecolor{CYAN}{cmyk}{1,0,0,0}
 \definecolor{MAGENTA}{cmyk}{0,1,0,0}
 \definecolor{YELLOW}{cmyk}{0,0,1,0}
}

\begin{document}
\title{Terahertz Strong-Field Physics without a Strong External Terahertz Field}
\normalsize

\author{Motoaki Bamba}
\thanks{Author to whom correspondence should be addressed}
\email[]{bamba@qi.mp.es.osaka-u.ac.jp}
\affiliation{Department of Materials Engineering Science, Osaka University, Toyonaka 560-8531, Japan}
\affiliation{PRESTO, Japan Science and Technology Agency, Kawaguchi 332-0012, Japan}

\author{Xinwei Li}
\affiliation{Department of Electrical and Computer Engineering, Rice University, Houston, Texas 77005, USA}

\author{Junichiro Kono}
\affiliation{Department of Electrical and Computer Engineering, Rice University, Houston, Texas 77005, USA}
\affiliation{Department of Physics and Astronomy, Rice University, Houston, Texas 77005, USA}
\affiliation{Department of Materials Science and NanoEngineering, Rice University, Houston, Texas 77005, USA}

\date{\today}

\def\Feion{\mathrm{Fe}^{3+}}
\def\Erion{\mathrm{Er}^{3+}}

%%%%%%%%%%%%%%%%%%%%%%%%
\begin{abstract}
Traditionally, strong-field physics explores phenomena in matter (atoms, molecules, and solids) driven by an extremely strong laser field nonperturbatively.  However, even in the complete absence of an external electromagnetic field, strong-field phenomena can arise when matter strongly couples with the zero-point field of the quantum vacuum state, i.e., fluctuating electromagnetic waves whose expectation value is zero.  Some of the most striking examples of this occur in a cavity setting, in which an ensemble of two-level atoms resonantly interacts with a single photonic mode of vacuum fields, producing vacuum Rabi splitting.  In particular, the nature of the matter-vacuum-field coupled system fundamentally changes when the coupling rate (equal to one half of the vacuum Rabi splitting) becomes comparable to, or larger than, the resonance frequency.  In this so-called ultrastrong coupling regime, a non-negligible number of photons exist in the ground state of the coupled system. Furthermore, the coupling rate can be cooperatively enhanced (via so-called Dicke cooperativity) when the matter is comprised of a large number of identical two-level particles, and a quantum phase transition is predicted to occur as the coupling rate reaches a critical value.  Low-energy electronic or magnetic transitions in many-body condensed matter systems with large dipole moments are ideally suited for searching for these predicted phenomena.  Here, we discuss two condensed matter systems that have shown cooperative ultrastrong interactions in the terahertz frequency range: a Landau-quantized two-dimensional electron gas interacting with high-quality-factor cavity photons, and an $\Erion$ spin ensemble interacting with $\Feion$ magnons in $\mathrm{ErFeO_3}$.
\end{abstract}

%\pacs{}

\maketitle
%%%%%%%%%%%%%%%%%%%%%%%%

\section{Introduction}

\newcommand{\bra}[1]{\langle #1|}
\newcommand{\ket}[1]{|#1\rangle}
\newcommand{\braket}[1]{\langle #1 \rangle}
\def\ketg{\ket{\text{g}}}

\def\dd{\mathrm{d}}
\def\ee{\mathrm{e}}
\def\ii{\mathrm{i}}
\def\rot{\bm{\nabla}\times}
\def\const{\text{const.}}
\def\ddt#1{\frac{\dd #1}{\dd t}}
\def\grad{\bm{\nabla}}

\def\diez{\varepsilon_0}
\def\die{\varepsilon}
\def\muz{\mu_0}

\def\wcyc{\omega_{\text{cyc}}}
\def\wcav{\omega_{\text{cav}}}

\def\oHH{\hat{\mathcal{H}}}
\def\oA{\hat{A}}
\def\oa{\hat{a}}
\def\oad{\hat{a}^{\dagger}}
\def\ob{\hat{b}}
\def\obd{\hat{b}^{\dagger}}
\def\oc{\hat{c}}
\def\ocd{\hat{c}^{\dagger}}
\def\oB{\hat{\beta}}
\def\oBd{\hat{\beta}^{\dagger}}

\def\oHo{\hat{H}_{\text{1e}}}
\def\oHLandau{\hat{H}_{\text{Landau}}}
\def\oHint{\hat{H}_{\text{int}}}
\def\oHdia{\hat{H}_{\text{dia}}}
\def\oHcavity{\hat{H}_{\text{cavity}}}
\def\oHcpl{\hat{H}_{\text{couple}}}
\def\ovA{\hat{\bm{A}}}
\def\ovr{\hat{\bm{r}}}
\def\ovp{\hat{\bm{p}}}
\def\ovpi{\hat{\bm{\pi}}}
\def\opi{\hat{\pi}}
\def\oAA{\hat{\mathcal{A}}}
\def\oPi{\hat{\varPi}}

\def\munit{\bm{1}}

\def\ml{l_{\text{m}}}
\def\dens{n_{\text{2DEG}}}
\def\area{S}
\def\len{L}
\def\NL{N_{\text{L}}}
\def\fluxq{\varPhi_{0}}
\def\diecav{\varepsilon_{\text{cav}}}
\def\diem{\varepsilon_{\text{2DEG}}}
\def\nL{n_{\text{L}}}
\def\rabi{g}

\def\vAz{\bm{A}_0}
\def\vBz{\bm{B}_0}

\def\Ac{\tilde{A}}

\def\thick{d}
\def\wp{\omega_{\text{plasma}}}

\def\rabit{\tilde{\rabi}}
\def\wA{\omega_{a}}
\def\wAt{\tilde{\omega}_{a}}
\def\wB{\omega_{b}}
\def\osigma{\hat{\sigma}}
\def\osigmad{\hat{\sigma}^{\dagger}}

\def\vr{\bm{r}}
\def\ovET{\hat{\bm{E}}_{\perp}}
\def\ovB{\hat{\bm{B}}}
\def\ovA{\hat{\bm{A}}}
\def\ovPi{\hat{\bm{\varPi}}}

\def\AA{D}
\def\Bstat{B_{\text{stat}}}
\def\vAstat{\bm{A}_{\text{stat}}}
\def\vBstat{\bm{B}_{\text{stat}}}

Matter in the presence of a strong electromagnetic (EM) field exhibits intriguing phenomena that cannot be understood by treating the field as a small perturbation~\cite{Bucksbaum90Book}. For example, the AC Stark effect, which occurs in the presence of a \emph{resonant} laser field, represents coupled light-matter states; a measure of the coupling strength is the Rabi energy, $\hbar \Omega_\text{Rabi} = d_{12}E$, where $d_{12}$ is the transition dipole moment and $E$ is the field strength. In the \emph{nonresonant} case, laser-driven matter exhibits other nonperturbative phenomena (e.g., above-threshold ionization and high-order harmonic generation), in which the ponderomotive potential $U_\text{p} = e^2E^2/(4m\omega^2)$, i.e., the cycle-averaged quiver energy of an electron in laser light with frequency $\omega$~\cite{Bucksbaum90Book,ChinetAl00PRL,SrivastavaetAl04PRL,Wegener05Book}, plays a critical role.  In both the resonant and nonresonant cases, different regimes of strong-field phenomena arise, depending on the normalized parameters $\hbar\Omega_\text{Rabi}/\hbar\omega$ and $U_\text{p}/\hbar\omega$, respectively.  Since these parameters scale as $1/\omega$ and $1/\omega^3$, respectively, for a given field strength $E$, it is inherently advantageous to use smaller laser frequencies to explore strong-field physics.  In particular, in condensed matter systems in the resonant case, there exist transitions with enormous $d_{12}$ in the terahertz (THz) frequency range, which makes it possible to explore uncharted regimes of strong-field physics even without a strong field; i.e., when frequency anticrossing induced by an interaction between the matter's transition and a vacuum EM field becomes comparable to, or larger than, the transition and photon frequencies, the so-called ultrastrong coupling (USC) regime arises~\cite{Forn-DiazetAl18arXiv,KockumetAl18arXiv}.

This article is concerned with electron systems that are ultrastrongly coupled with photons in a cavity (or bosons with a resonant frequency, in more general, including magnons and phonons in long-wavelength limit). There are a variety of theoretically predicted many-body cavity quantum electrodynamics (QED) effects, states, and phases in the USC regime, which also offer possibilities for constructing unique protocols for ultrafast gates and ultrasecure state preparation for quantum information processing~\cite{Forn-DiazetAl18arXiv,KockumetAl18arXiv}.  One of the most exciting aspects of cavity QED physics in the USC regime is the fact that the ``light field'' that the matter strongly couples with is not an external laser field but the vacuum cavity field, whose amplitude's expectation value is zero. This fact distinctly differentiates USC physics from ordinary nonlinear optical phenomena, which are induced by a strong {\em external} field and thus inevitably involve excited and/or nonequilibrium states of matter.  On the contrary, USC phenomena exclusively reflect the properties of \emph{the ground state of the matter--vacuum-field hybridized system in equilibrium}.  This new ground state, represented by a matter--vacuum-field entangled wavefunction~\cite{CiutietAl05PRB,AshhabNori10PRA,FelicettietAl15SR}, has characteristics that neither the original matter ground state (which can be metallic or insulating) nor the usual vacuum (which consists of fluctuating EM fields) possesses.

%In a cavity QED system, there are four quantities that jointly characterize different light-matter coupling regimes: $\omega_0$, $g$, $\kappa$, and $\gamma$. Here, $\omega_0$ is the cavity mode photon frequency.  The parameter $g$ is the coupling constant (or rate), with 2$g$ being the vacuum Rabi splitting (VRS) between the two normal modes, the lower polariton (LP) and upper polariton (UP), of the coupled system. The parameter $\kappa$ is the photon decay rate of the cavity; $\tau_\mathrm{cav}$ $=$ $\kappa^{-1}$ is the photon lifetime of the cavity, and the cavity quality factor $Q$ $=$ $\omega_0 \tau_\mathrm{cav}$.  The parameter $\gamma$ is the nonresonant matter decay rate, which is usually the decoherence rate in the case of solids.  Strong coupling (\textbf{SC}) is achieved when the VRS, $2g$, is much larger than the linewidth, ($\kappa + \gamma)/2$, \textbf{USC} is achieved when $g$ becomes a considerable fraction of $\omega_{0}$, and some authors define the deep strong coupling (\textbf{DSC}) regime as $g/\omega_0 > 1$~\cite{CasanovaetAl10PRL}. The two standard figures of merit to measure the coupling strength are $C$ $\equiv$ $4g^2/(\kappa\gamma)$ and $g/\omega_0$; here, $C$ is called the {\em cooperativity} parameter~\cite{BonifacioLugiato82Book}, which is also the determining factor for the onset of optical bistability through resonant absorption saturation.  

%%%%%%%%%%%%%%%%%%%%%%%%% figure 1 %%%%%%%%%%%%%%%%%%%%%
\begin{figure*}
\begin{center}
\begin{tabular}{c}
\includegraphics[width=0.9\linewidth]{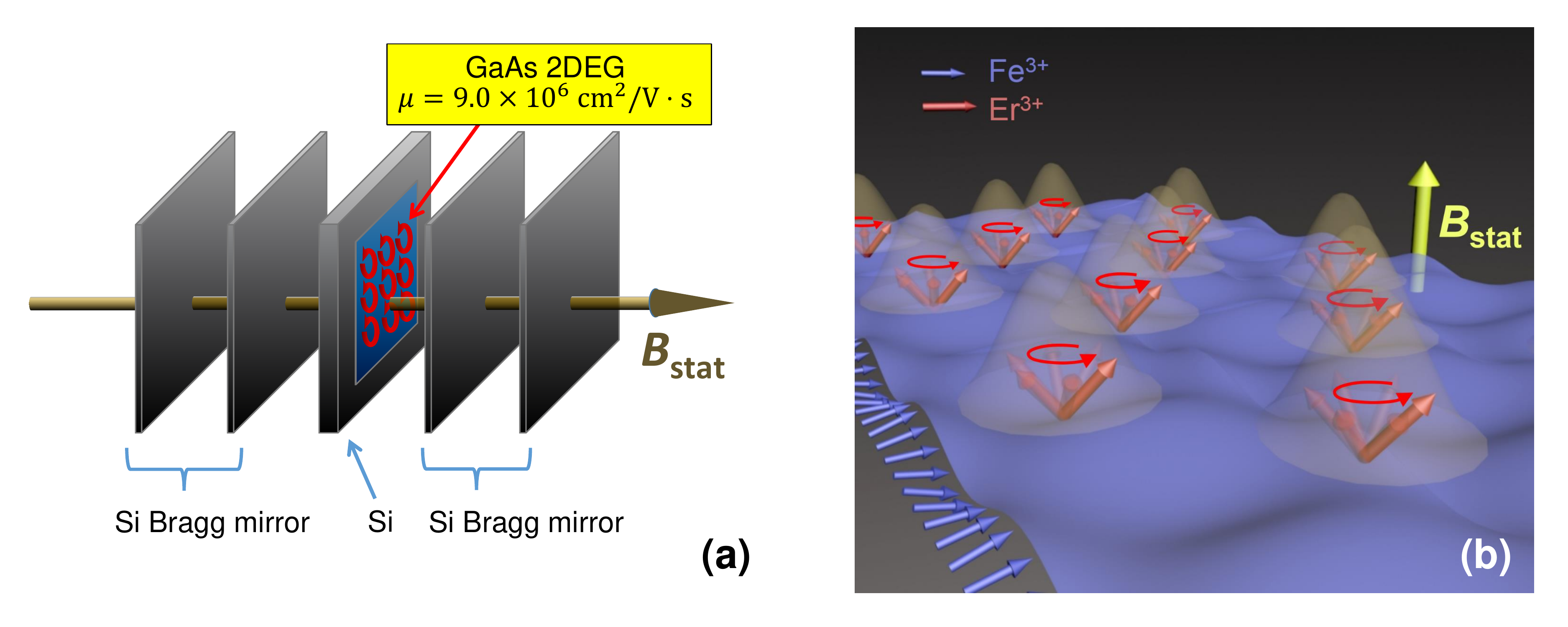}
\end{tabular}
\end{center}
\caption{(a)~Ultrastrong coupling between the cyclotron resonance of a two-dimensional electron gas and THz photons in a cavity~\cite{ZhangetAl16NP}. A vacuum Bloch--Siegert shift was experimentally observed~\cite{LietAl18NP}. (b)~Ultrastrong coupling between $\mathrm{Er}^{3+}$ spins and $\mathrm{Fe}^{3+}$ magnons in a crystal of the canted antiferromagnet $\mathrm{ErFeO_3}$. The Dicke cooperatively was observed in their interaction strength as a function of $\Erion$ density~\cite{LietAl18Science}.\label{intro}}
\end{figure*}
%%%%%%%%%%%%%%%%%%%%%%%%%%%%%%%%%%%%%%%%%%%%%%%%%%%%%%%%

Specifically, in this article, we describe two systems shown in Fig.\,\ref{intro}: (a)~an ultrahigh-mobility two-dimensional electron gas (2DEG) in GaAs quantum wells placed inside a high-quality-factor($Q$) THz cavity~\cite{ZhangetAl16NP,LietAl18NP} and (b)~an ensemble of $\mathrm{Er}^{3+}$ spins interacting with $\mathrm{Fe}^{3+}$ magnons in a $\mathrm{ErFeO_3}$ crystal~\cite{LietAl18Science}.  In both systems, we have observed USC --- between the 2DEG and cavity photons in the former, and between electron spins and magnons in the latter. In order to describe the novel strong-field phenomena we observed in these systems, in Section \ref{sec:Hamiltonian} below, we present various types of Hamiltonians with different degrees of approximations appropriate for describing light--matter interactions in different coupling strength regimes.  In the subsequent sections, Sections \ref{sec:Landau} and \ref{sec:ErFeO3}, we present more detailed models and results on the two systems. Finally, we summarize our findings and conclusions in Section \ref{sec:summary}.

\section{Light--Matter Interaction Hamiltonians} \label{sec:Hamiltonian}

In the presence of nonperturbative, or ultrastrong, coupling between light and matter, in general, we can no longer use the rotating-wave approximation (RWA) for describing light-matter interaction phenomena. Further, we can no longer neglect the ponderomotive energy $U_\text{p}$~\cite{Bucksbaum90Book,ChinetAl00PRL,SrivastavaetAl04PRL,Wegener05Book}. Even when the light is quite weak or even in the vacuum level, these kinds of approximations also fail when materials ultrastrongly interact with the EM fields~\cite{CiutietAl05PRB,Forn-DiazetAl18arXiv,KockumetAl18arXiv}, i.e., when the interaction strength $\rabi$ (which is equal to one half of the vacuum Rabi splitting) becomes comparable to, or larger than, the photon frequency $\wA$ and the transition frequency $\wB$ of the material.

In the case of $\rabi\ll\wA,\wB$, a system consisting of a photon mode with a resonance frequency $\wA$ and an ensemble of $N$ two-level atoms with a transition frequency $\wB$ can be described by the following Hamiltonian:
\begin{widetext}
\begin{align} \label{eq:oH_TC} % !!!!!!!!!!!!!!!!!!!!!!!!!!!!!!!!!!!!!!!!!!
\oHH_{\text{Tavis--Cummings}} &= \hbar\wA\oad\oa
+ \hbar\wB\sum_{i=1}^N\frac{\osigma_{i,z}}{2} 
 + \frac{\ii\hbar\rabi}{\sqrt{N}}\sum_{i=1}^N
  \left(
    \osigmad_i\oa - \oad\osigma_i
  \right).
\end{align}
\end{widetext}
In this paper, we call this the Tavis--Cummings Hamiltonian, following Ref.~\citenum{KockumetAl18arXiv}. Here, $\oa$ ($\oad$) is the annihilation (creation) operator of a photon. $\osigma_{i,z} = \ket{\uparrow_i}\bra{\uparrow_i}-\ket{\downarrow_i}\bra{\downarrow_i}$ is one of the Pauli operators representing the state of the $i$-th two-level atom, and $\osigma_i=\ket{\downarrow_i}\bra{\uparrow_i}$ is its lowering operator from the excited state $\ket{\uparrow_i}$ to the ground state $\ket{\downarrow_i}$. The last two terms in the Hamiltonian, $\osigmad\oa$ and $\oad\osigma_i$, represent, respectively, the excitation of the atom through absorption of a photon and the relaxation of the atom through emission of a photon. $\rabi/\sqrt{N}$ is the interaction strength per atom, while $\rabi$ is that for the ensemble (collective excitation) of the atoms. In other words, the interaction strength $\rabi$ scales as $\sqrt{N}$ with the increase in the number $N$ of atoms (precisely speaking, $\rabi$ is proportional to the density of atoms beyond the long-wavelength approximation).

As discussed by R.~H.~Dicke in 1954~\cite{Dicke54PR}, the above Hamiltonian is usually derived from the minimal-coupling Hamiltonian (in the Coulomb gauge)~\cite{Cohen-TannoudjietAl89Book}:
\begin{widetext}
\begin{equation} \label{eq:oH_min} % !!!!!!!!!!!!!!!!!!!!!!!!!!!!!!!!!!!!!!!!!!
\oHH_{\text{min}}
= \int\dd\vr \left[ \frac{\diez\ovET(\vr)^2}{2} + \frac{\ovB(\vr)^2}{2\muz} \right]
+ \sum_{i=1}^N \left[ \frac{[\ovp_i + e\ovA(\ovr_i)]^2}{2m} + V(\{\ovr_i\}) \right].
\end{equation}
\end{widetext}
Here, $\ovET(\vr) = -\ovPi(\vr)/\diez$ and $\ovB(\vr) = \rot\ovA(\vr)$ are the transverse electric field and magnetic flux density, respectively, expressed by the vector potential $\ovA(\vr)$ and its conjugate momentum $\ovPi(\vr)$ satisfying $[\ovA(\vr), \ovPi(\vr')] = \ii\hbar\bm{\delta}_{\perp}(\vr-\vr')$, where $\bm{\delta}_{\perp}(\vr)$ is the transverse dyadic delta function~\cite{Cohen-TannoudjietAl89Book}. $\ovr_i$ and $\ovp_i$ are the position and momentum of the $i$-th electron with charge $-e$ and mass $m$, satisfying $[\ovr_i,\ovp_{j}]=\ii\hbar\delta_{i,j}\bm{1}$. The last term $V(\{\ovr\})$ represents the electrons' one-body potential and many-body interactions in general. Here, we assume that, in this one-body potential, the oscillator strength is concentrated on a transition between the atomic ground state and an excited state with energy difference $\hbar\wB$. We also assume that the atoms are well isolated from each other and $V(\{\ovr\})$ does not give many-body interactions. Further, we assume that the photons are confined in a cavity (embedding the atoms) with a resonance frequency $\wA$, and we focus only on one of the transverse directions of the polarization. Under these assumptions, in the long-wavelength approximation, $\oHH_{\text{min}}$ can be rewritten as~\cite{Dicke54PR,RzazewskietAl75PRL,BambaOgawa14PRA2}
\begin{widetext}
\begin{align} \label{eq:oH_Dicke} % !!!!!!!!!!!!!!!!!!!!!!!!!!!!!!!!!!!!!!!!!!
\oHH_{\text{Dicke}} & = \hbar\wA\oad\oa
+ \hbar\wB\sum_{i=1}^N\frac{\osigma_{i,z}}{2}
+ \frac{\ii\hbar\rabi}{\sqrt{N}}
  \left(\oad + \oa\right)\sum_{i=1}^N\left(\osigmad_i - \osigma_i\right)
+ \hbar\AA \left(\oad + \oa\right)^2.
\end{align}
\end{widetext}
In this paper, we call this the Dicke Hamiltonian. The last term is called the $A^2$, diamagnetic, or quadratic term. Its coefficient is usually derived as $\AA=\rabi^2/\wB$ under the above assumptions. Here, $\rabi^2$ is proportional to the oscillator strength or $d_{12}{}^2$ of the two-level atom, the atom density, and the vacuum fluctuation $\braket{\oA\oA}$ of the vector potential (in contrast to the ponderomotive energy~\cite{Bucksbaum90Book,ChinetAl00PRL,SrivastavaetAl04PRL,Wegener05Book}, which is proportional to the intensity of light). 

In the limit of $\rabi\ll\wA,\wB$, we can use the RWA (eliminating the counter-rotating terms $\oad\osigmad_i$ and $\oa\osigma_i$) and neglect the last term in Eq.~\eqref{eq:oH_Dicke}. Therefore, in that limit, $\oHH_{\text{Dicke}} \approx \oHH_{\text{Tavis--Cummings}}$. However, in the USC regime, where $\rabi\gtrsim\wA,\wB$, these approximations cannot be justified. Due to the presence of the counter-rotating terms, the expectation value $\braket{\oad\oa}$ of the photon number becomes nonzero even in the ground state of the total system, and these photons are called virtual photons~\cite{CiutietAl05PRB}. If the coefficient of the $A^2$ term satisfies $\AA < \rabi^2/\wB$, we get a thermal second-order phase transition called the superradiant phase transition (SRPT)~\cite{HeppLieb73AP,WangHioe73PRA,NatafCiuti10NC}, where the expectation value $\braket{\oa}$ of the EM field spontaneously becomes nonzero (not temporally oscillating) below a critical temperature in thermal equilibrium. In particular, when $\AA=0$, the SRPT occurs if the interaction strength satisfies $\rabi^2>\wA\wB/4$ (i.e., the USC regime). However, considering the sum rule of the oscillator strengths, we usually get $\AA \ge \rabi^2/\wB$, which forbids the SRPT~\cite{RzazewskietAl75PRL,FN1}.
%
%\footnote{\label{fn:1}We can eliminate the $A^2$ term by a Bogoliubov transformation as $\oHH_{\text{Dicke}}=\hbar\wAt\oa^{\prime\dagger}\oa'+\hbar\wB\sum_{i=1}^N(\osigma_{i,z}/2) + (\ii\hbar\rabit/\sqrt{N})(\oa^{\prime\dagger}+\oa')\sum_{i=1}^N(\osigmad_i-\osigma_i) + \const$, where $\wAt \equiv \sqrt{\wA{}^2+4\AA\wA}$, $\rabit\equiv \rabi\sqrt{\wA/\wAt}$, and the photon operator is renormalized as $\oa'\equiv [(\wAt+\wA)\oa+(\wAt-\wA)\oad]/(2\sqrt{\wAt\wA})$. However, if $\AA\ge\rabi^2/\wB$ is satisfied, we always get $\rabit^2 < \wAt\wB/4$, and thus, the SRPT does not occur.} 
A more general no-go theorem of the SRPT was also proposed in the minimal-coupling Hamiltonian, Eq.~\eqref{eq:oH_min}~\cite{Bialynicki-BirulaRzazewski79PRA,GawedzkiRzazewski81PRA}. Counter examples against that no-go theorem are still being discussed even in the current literature~\cite{GriesseretAl16PRA,MazzaGeorges19PRL}. However, the thermal SRPT has not yet been realized experimentally, while its nonequilibrium analogue has been demonstrated in cold atoms~\cite{BaumannetAl10Nature}, and a thermal-equilibrium analogue has been theoretically proposed in an artificial system composed of a superconducting circuit~\cite{BambaetAl16PRL}.

In most experiments~\cite{Forn-DiazetAl18arXiv}, the interaction strength $\rabi$ and the coefficient $\AA$ of the $A^2$ term can be estimated through linear optical responses (absorption, transmission, or reflection spectra). In such cases, we can bosonize the ensemble of the atoms (through the lowest-level Holstein--Primakoff transformation), i.e., $\sum_{i=1}^N(\osigma_{i,z}/2)\to\obd\ob+\const$ and $\sum_{i=1}^N\osigma_i/\sqrt{N}\to\ob$, where $\ob$ ($\obd$) is the annihilation (creation) operator of the collective excitation of the atoms. Then, the Dicke Hamiltonian is approximated as
\begin{widetext}
\begin{align} \label{eq:oH_Hopfield} % !!!!!!!!!!!!!!!!!!!!!!!!!!!!!!!!!!!!!!!!!!
\oHH_{\text{Hopfield}} & = \hbar\wA\oad\oa
+ \hbar\wB\obd\ob
+ \ii\hbar\rabi  \left(\oad + \oa\right)\left(\obd - \ob\right)
+ \hbar\AA\left(\oad + \oa\right)^2.
\end{align}
\end{widetext}
In this paper, we call this the Hopfield Hamiltonian. This approximation is justified in the weak excitation limit. The virtual photons appear also in the ground state of this Hamiltonian due to the counter-rotating terms. The SRPT cannot be described by the Hopfield Hamiltonian. Instead, we find an instability of the normal ground state (showing $\braket{\oa}=\braket{\osigma_i}=0$) under the same condition as that for the SRPT ($\rabi^2>\wA\wB/4$ for $\AA=0$)~\cite{FN2}.
%
%\footnote{We can find that, in the case of $\rabi^2>\wA\wB/4$ and $\AA=0$, an eigenfrequency of the Hopfield Hamiltonian in Eq.~\eqref{eq:oH_Hopfield} becomes a pure imaginary value, which indicates the instability of the normal ground state. This is because the quartic part of the potential of the atomic collective excitation is neglected in $\oHH_{\text{Hopfield}}$, although it is essential for describing the double-well potential of the second-order phase transition in the Landau theory. When we consider the higher-order terms, the SRPT can be reproduced in the Hamiltonian after the Holstein-Primakoff transformation.\cite{EmaryBrandes03PRL,EmaryBrandes03PRE}}

%\cite{EmaryBrandes03PRL,EmaryBrandes03PRE}
For experimentally discussing the virtual photons and the existence of the SRPT, it is necessary to quantitatively evaluate the contributions of the counter-rotating terms and the $A^2$ term. In one of our experiments~\cite{LietAl18NP}, these contributions were evaluated separately by using a high-$Q$ THz cavity integrated with an ultrahigh-mobility 2DEG. As shown in Fig.\,\ref{intro}(a), the THz cavity had a pair of silicon Bragg reflectors, each of which consisted of alternating layers of silicon and vacuum, and a defect thick silicon layer in the middle. The 2DEG, made of modulation-doped GaAs quantum wells grown by molecular-beam epitaxy, was attached to one of the surfaces of the defect layer; its position also overlapped with the position where the electric field reaches maximum at resonance. Under an external static magnetic flux density $\Bstat$, the cyclotron motion of the 2DEG is excited resonantly by a THz wave with corotating circular polarization, which showed an anticrossing with the cyclotron resonance (CR) as we will see in Fig.~\ref{VBS}. On the other hand, we found a frequency shift of the cavity mode with a circular polarization counter-rotating against the cyclotron motion (see the shaded area in Fig.~\ref{VBS}). This shift is interpreted as the vacuum counterpart of the Bloch--Siegert (BS) shift~\cite{BlochSiegert40PR}, which usually appears due to an electron's counter-rotating coupling with a strong light field. This vacuum BS shift was observed even when the average photon number was close to zero owing to the ultrastrong coupling, high-$Q$ cavity, and high-mobility 2DEG in our system. In order to describe the behavior of the resonance frequencies depending on the circular polarization, we need to consider explicitly the polarization degree of freedom of photons coupled with the CR, where the time reversal symmetry is broken under the static magnetic flux density $\Bstat$. In contrast to the Hopfield Hamiltonian, Eq.~\eqref{eq:oH_Hopfield}, consisting of a single photon mode, the Hamiltonian in our system is expressed as
\begin{widetext}
\begin{align} \label{eq:oH_Landau} % !!!!!!!!!!!!!!!!!!!!!!!!!!!!!!!!!!!!!!!!!!
\oHH_{\text{Landau}}
& = \sum_{\xi=\pm}\hbar\wA\oad_{\xi}\oa_{\xi} + \hbar\wB\obd\ob
+ \ii\hbar\rabi
  \left[
    \obd(\oa_{+} + \oad_{-})
  - (\oa_{-} + \oad_{+})\ob
  \right]
+ \frac{\hbar\rabi{}^2}{\wB}
  (\oa_{-}+\oad_{+})(\oa_{+}+\oad_{-}).
\end{align}
\end{widetext}
Here, $\oa_+$ ($\oa_-$) is the annihilation operator of a photon with a circular polarization corotating (counter-rotating) against the cyclotron motion of the 2DEG in $\Bstat$. $\ob$ is the annihilation operator of the collective excitation of the 2DEG between the Landau levels with a transition frequency $\wB = e|\Bstat|/m$. $\oa_+$ interacts with $\ob$ in the corotating manner as $\ii\hbar\rabi(\obd\oa_+-\oad_+\ob)$, while $\oa_-$ interacts with $\ob$ in the counter-rotating manner as $\ii\hbar\rabi(\obd\oad_--\oa_-\ob)$. Because of this difference, we could clearly and quantitatively differentiate between the contributions of the counter-rotating terms (vacuum BS shift) and the $A^2$ terms, which will be discussed in Sec.~\ref{sec:Landau} in addition to the derivation of $\oHH_{\text{Landau}}$.

While the presence of a thermal SRPT in the minimal-coupling Hamiltonian (derived from the Maxwell equations and Newton's equation of charged particles feeling the Lorentz force) in Eq.~\eqref{eq:oH_min} is still under debate~\cite{Bialynicki-BirulaRzazewski79PRA,GawedzkiRzazewski81PRA,GriesseretAl16PRA,MazzaGeorges19PRL}, there remains a possibility for realizing a thermal SRPT in systems where spins, instead of charges, interact with the EM fields, as pointed out by Knight {\it et al.} in 1978~\cite{KnightetAl78PRA}. A hint of such a possibility lies in certain magnetic phase transitions, which are caused by interactions between two species of spins within the same material. In particular, we focus on magnetic materials that can be modeled by the Dicke Hamiltonian under the replacement of photons with magnons in one of the spin species. If the magnetic phase transitions in such systems can be interpreted as magnon analogues of the SRPT, they will provide clues toward realizing the original photon SRPT. Recently, we have experimentally obtained evidence for ultrastrong and cooperative interactions between an ensemble of $\mathrm{Er}^{3+}$ ions and a magnon mode of the $\mathrm{Fe}^{3+}$ lattice in $\mathrm{ErFeO_3}$ crystals. As shown in Fig.\,\ref{intro}(b), ordered $\mathrm{Fe}^{3+}$ spins form a magnon wave, while the electron paramagnetic resonance (EPR) of the $\mathrm{Er}^{3+}$ ions is interpreted as an atomic transition interacting with the magnon wave. Eventually, we can reduce a spin model of $\mathrm{ErFeO_3}$ into the Dicke Hamiltonian (with $\AA=0$) as
\begin{widetext}
\begin{align} \label{eq:oH_ErFeO3} % !!!!!!!!!!!!!!!!!!!!!!!!!!!!!!!!!!!!!!!!!!
\oHH_{\mathrm{ErFeO_3}}
& = \hbar\wA\oad\oa + \hbar\wB\sum_{i=1}^N\frac{\osigma_{i,z}}{2}
+ \frac{\ii\hbar\rabi}{\sqrt{N}}\sum_{i=1}^N
  \left(\oad+\oa\right)\left(\osigmad_i-\osigma_i\right).
\end{align}
\end{widetext}
Here, $\oa$ is the annihilation operator of a magnon in the quasiferromagnetic (qFM) mode in the $\mathrm{Fe}^{3+}$ lattice. $\osigma_i$ is the lowering operator between the lowest two $\mathrm{Er}^{3+}$ levels in a static magnetic field. As we will see in Fig.~\ref{ErFO}, we experimentally found an anticrossing between the $\mathrm{Er}^{3+}$ EPR and the qFM magnon mode, and the vacuum Rabi splitting ($2\rabi$) reached a considerable fraction of the EPR and magnon frequencies.
%but also shows characteristic scaling behavior with the $\mathrm{Er}^{3+}$ density as predicted by the Dicke model.
Further, by replacing $\mathrm{Er}^{3+}$ with $\mathrm{Y}^{3+}$ through chemical doping, we also found that the vacuum Rabi splitting $2\rabi$ was proportional to the square root of the $\mathrm{Er}^{3+}$ density. These are discussed in Sec.~\ref{sec:ErFeO3} in addition to the derivation of the Dicke Hamiltonian in Eq.~\eqref{eq:oH_ErFeO3} from the spin model of $\mathrm{ErFeO_3}$.

\section{Cyclotron resonance interacting with THz photons} \label{sec:Landau}
\def\dd{\mathrm{d}}
\def\ee{\mathrm{e}}
\def\ii{\mathrm{i}}
\def\rot{\bm{\nabla}\times}
\def\const{\text{const.}}
\def\ddt#1{\frac{\dd #1}{\dd t}}

\def\diez{\varepsilon_0}
\def\die{\varepsilon}

\def\wcyc{\omega_{\text{cyc}}}
\def\wcav{\omega_{\text{cav}}}

\def\oH{\hat{H}}
\def\oA{\hat{A}}
\def\oa{\hat{a}}
\def\oad{\hat{a}^{\dagger}}
\def\ob{\hat{b}}
\def\obd{\hat{b}^{\dagger}}
\def\oc{\hat{c}}
\def\ocd{\hat{c}^{\dagger}}
\def\oB{\hat{\beta}}
\def\oBd{\hat{\beta}^{\dagger}}

\def\oHo{\hat{H}_{\text{1e}}}
\def\oHLandau{\hat{H}_{\text{Landau}}}
\def\oHint{\hat{H}_{\text{int}}}
\def\oHdia{\hat{H}_{\text{dia}}}
\def\oHcavity{\hat{H}_{\text{cavity}}}
\def\oHcpl{\hat{H}_{\text{couple}}}
\def\ovA{\hat{\bm{A}}}
\def\ovr{\hat{\bm{r}}}
\def\ovp{\hat{\bm{p}}}
\def\ovpi{\hat{\bm{\pi}}}
\def\opi{\hat{\pi}}
\def\oAA{\hat{\mathcal{A}}}
\def\oPi{\hat{\varPi}}

\def\munit{\bm{1}}

\def\ml{l_{\text{m}}}
\def\dens{n_{\text{2DEG}}}
\def\area{S}
\def\len{L}
\def\NL{N_{\text{L}}}
\def\fluxq{\varPhi_{0}}
\def\diecav{\varepsilon_{\text{cav}}}
\def\diem{\varepsilon_{\text{2DEG}}}
\def\nL{n_{\text{L}}}

\def\vAz{\bm{A}_0}
\def\vBz{\bm{B}_0}

\def\Ac{\tilde{A}}

\def\thick{d}
\def\wp{\omega_{\text{plasma}}}

\def\oHH{\hat{\mathcal{H}}}
\def\oHHem{\oHH_{\text{EM}}}
\def\oHHel{\oHH_{\text{2DEG}}}
\def\HH{\mathcal{H}}
\def\HHem{\HH_{\text{em}}}
\def\muz{\mu_0}
\def\ddt#1{\frac{\partial #1}{\partial t}}
\def\ddz#1{\frac{\partial #1}{\partial z}}
\def\ddzz#1{\frac{\partial^2 #1}{\partial z^2}}
\def\ddtt#1{\frac{\partial^2 #1}{\partial t^2}}
\def\oE{\hat{E}}
\def\oB{\hat{B}}
\def\vE{\bm{E}}
\def\vB{\bm{B}}
\def\vA{\bm{A}}
\def\pos{z_{\text{2DEG}}}
\def\rabiz{\bar{\rabi}}
\def\rabit{\tilde{\rabi}}
\def\diebg{\die_{\text{bg}}}
\def\diecav{\varepsilon_{\text{cav}}}
\def\Leff{L_{\text{eff}}}

\def\GHz{\mathrm{GHz}}
\def\micron{\mathrm{\mu{}m}}

As shown in Fig.~\ref{intro}(a), we consider a planar cavity embedding a 2DEG parallel to the cavity mirrors. Generalizing the minimal-coupling Hamiltonian in Eq.~\eqref{eq:oH_min}, we explicitly consider the cavity mirror structure described by a position-dependent relative dielectric constant $\diecav(z)$, where $z$ is the direction perpendicular to the mirrors and the 2DEG. For simplicity, we focus only on the EM wave without an in-plane wavenumber. The total Hamiltonian is expressed as
\begin{equation}
\oHH_{\text{Landau}} = \oHHem + \oHHel.
\end{equation}
Following the quantization of the EM wave in an inhomogeneous dielectric medium in Ref.~\citenum{GlauberLewenstein91PRA}, we describe the Hamiltonian of the EM wave (excluding the external static field) as
%\begin{widetext}
\begin{align}
\oHHem
& = \sum_{\xi=x,y} \int\dd z \left[ \frac{\diez\diecav(z)\oE_\xi(z)^2}{2} + \frac{\oB_\xi(z)^2}{2\muz} \right] \nonumber \\& = \sum_{\xi=x,y} \int\dd z \left[ \frac{\oPi_\xi(z)^2}{2\diez\diecav(z)} + \frac{1}{2\muz}\left(\ddz{}\oA_\xi(z)\right)^2\right].
\end{align}
%\end{widetext}
The conjugate momentum of the vector potential $\oA_{\xi}(z)$
corresponds to the electric displacement field in the cavity medium (but without the 2DEG) as
$\oPi_{\xi}(z) = - \diez\diecav(z)\oE_{\xi}(z)$. These operators satisfy $[\oA_{\xi}(z), \oPi_{\xi'}(z')] = \ii\hbar\delta_{\xi,\xi'}\delta(z-z')/\area$, where $\area$ is the area of the $x$--$y$ plane. On the other hand, the kinetic energy of the 2DEG is expressed as
\begin{align}
\oHHel & = \sum_{i=1}^N \sum_{\xi=x,y}\frac{[\opi_{i,\xi}+e\oA_\xi(\pos)]^2}{2m}.
\end{align}
Here, $\ovpi_i\equiv\ovp_i+e\vAstat$ and $\vAstat$ gives the external static magnetic flux density $\vBstat=\rot\vAstat$. $\pos$ is the position of the 2DEG inside the cavity. $N = \dens\area$ is the number of electrons for surface density $\dens$ of the 2DEG. We do not consider electron-electron Coulomb interactions, since they do not affect the linear optical response of the 2DEG due to Kohn's theorem~\cite{Kohn61PR}.

Let us first rewrite the EM Hamiltonian $\oHHem$ in terms of the annihilation and creation operators of the photons. From the Heisenberg or Hamilton's equations (giving the Maxwell equations) derived from $\oHHem$, we get a wave equation for the vector potential $\oA_{\xi}(z)$ as
\begin{equation}
\ddzz{}\oA_\xi(z,t) - \frac{\diecav(z)}{c^2}\ddtt{}\oA_\xi(z,t) = 0.
\end{equation}
Following the quantization procedure in Ref.~\citenum{GlauberLewenstein91PRA}, by replacing $\oA_\xi(z,t)$ with $f_k(z,t)/\sqrt{\diecav(z)}$ and by performing the temporal Fourier transformation, we rewrite the wave equation as
\begin{equation}
\frac{1}{\sqrt{\diecav(z)}}\ddzz{}\frac{f_k(z)}{\sqrt{\diecav(z)}} + \frac{\omega_k{}^2}{c^2} f_k(z) = 0.
\end{equation}
Solving this equation, we can determine the eigenfrequencies $\{\omega_k\}$ and eigenfunctions $\{f_k(z)\}$ of the EM wave in the cavity. $k$ is the mode index. The wavefunctions are normalized as $\int\dd z\ f_k(z) f_{k'}(z) = \delta_{k,k'}$. They also satisfy the completeness as $\sum_k f_k(z) f_{k}(z') = \delta(z-z')$. From the complete set of these eigenmodes, the operators of the vector potential and its conjugate momentum are described as
\begin{align}
\oA_\xi(z) & = \sum_k \sqrt{\frac{\hbar}{2\diez\diecav(z)\omega_k\area}} f_k(z) \left( \oad_{k,\xi} + \oa_{k,\xi} \right), \\
\oPi_\xi(z) & = \sum_k \ii \sqrt{\frac{\diez\diecav(z)\hbar\omega_k}{2\area}} f_k(z) \left( \oad_{k,\xi} - \oa_{k,\xi} \right).
\label{eq:oPi}
\end{align}
Here, $\oa_{k,\xi}$ is the annihilation operator of a photon in the $k$-th mode with a linear polarization in the $\xi=x,y$ direction, satisfying $[\oa_{k,\xi}, \oad_{k',\xi'}] = \delta_{k,k'}\delta_{\xi,\xi'}$. The Hamiltonian of the EM wave is then rewritten as
\begin{equation}
\oHHem = \sum_{k}\sum_{\xi=x,y}\hbar\omega_k\left( \oad_{k,\xi}\oa_{k,\xi} + \frac{1}{2} \right).
\end{equation}
In general, there are not only the cavity modes (spatially localized modes)
but also continuous modes (transmission modes).

On the other hand, the 2DEG shows the CR with a frequency $\wcyc = e|\Bstat|/m$. Introducing the lowering operator $\oc_i\equiv(\opi_{i,y}+\ii\opi_{i,x})/\sqrt{2m\hbar\wcyc}$
between the Landau levels satisfying $[\oc_i, \ocd_j] = \delta_{i,j}$~\cite{ElvangPolchinski03CRP}, we rewrite the kinetic energy of the 2DEG as
\begin{widetext}
\begin{align}
\oHHel
& = \sum_{i=1}^N\hbar\wcyc\left( \ocd_i\oc_i + \frac{1}{2} \right)
+ \ii \sqrt{\frac{\hbar\wcyc e^2}{m}}
  \sum_{i=1}^N[\ocd_i\oA_+(\pos) - \oA_-(\pos)\oc_i]
+ \frac{Ne^2}{2m}\ovA(\pos)^2.
\end{align}
\end{widetext}
The last term is the $A^2$ term. The second term contains the lowering and raising processes involved with the non-Hermitian vector potential
%\begin{widetext}
\begin{align}
\oA_{\pm}(z) &\equiv \frac{\oA_x(z)\mp\ii\oA_y(z)}{\sqrt{2}} \nonumber
\\ &= \sum_k \sqrt{\frac{\hbar}{2\diez\diecav(z)\omega_k\area}} f_k(z)\left(  \oad_{k,\mp} + \oa_{k,\pm} \right),
\end{align}
%\end{widetext}
where the annihilation operator of the $\pm$ circularly polarized photon in the $k$-th mode is defined as
\begin{equation}
\oa_{k,\pm} \equiv \frac{\oa_{k,x} \mp \ii \oa_{k,y}}{\sqrt{2}}.
\end{equation}
Introducing the bosonic operator of a collective excitation of the 2DEG coherently interacting with the EM wave
\begin{equation}
\ob \equiv \frac{1}{\sqrt{N}}\sum_{i=1}^N\oc_i,
\end{equation}
we can rewrite the interaction term as
\begin{align}
\ii \sqrt{\frac{\hbar\wcyc e^2}{m}}
  \sum_{i=1}^N\left[ \ocd_i\oA_+(\pos) - \oA_-(\pos)\oc_j\right] \nonumber \\
= \sum_k \ii\hbar\rabi_k
    \left[ \obd (\oa_{k,+}+\oad_{k,-}) - (\oa_{k,-}+\oad_{k,+}) \ob \right],
\end{align}
where the interaction strength for mode $k$ is expressed as
\begin{equation}
\rabi_k \equiv \sqrt{\frac{e^2 \wcyc \dens}{2\diez\diecav(\pos)m\omega_k}} f_k(\pos).
\end{equation}
On the other hand, the $A^2$ term is rewritten as
\begin{align}
\frac{Ne^2}{2m}\ovA^2
= \sum_{k,k'} \frac{\hbar\rabi_k\rabi_{k'}}{\wcyc}
    (\oa_{k,-}+\oad_{k,+})(\oa_{k',+}+\oad_{k',-}).
\end{align}
Then, when we focus only on the CR collective excitation described by $\ob$ and one cavity mode with resonance frequency $\wA = \omega_k$, the total Hamiltonian is finally expressed as Eq.~\eqref{eq:oH_Landau}.

%%%%%%%%%%%%%%%%%%%%%%%%% figure 2 %%%%%%%%%%%%%%%%%%%%%
\begin{figure*}
\begin{center}
\begin{tabular}{c}
\includegraphics[width=\linewidth]{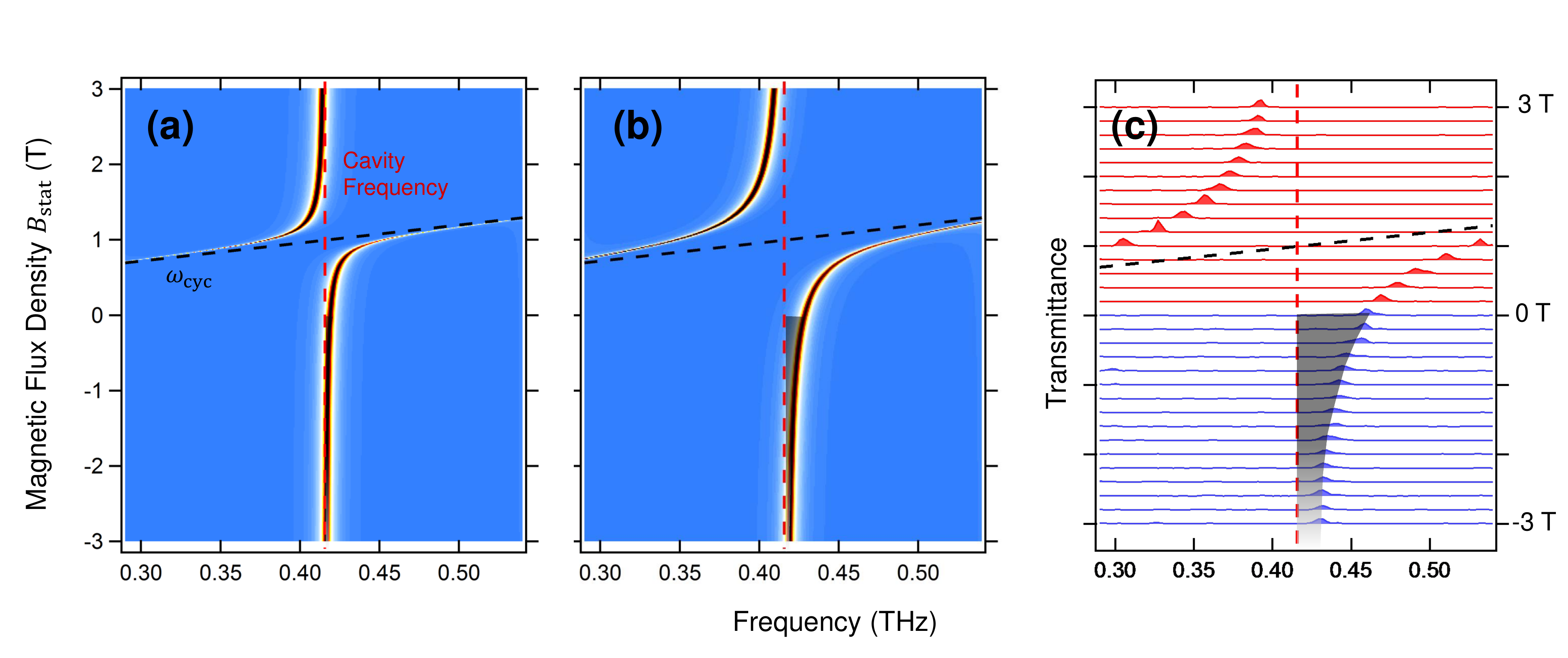}
\end{tabular}
\end{center}
\caption{Numerically calculated THz transmission for the CR--cavity system for interaction strength (a) $\rabi/2\pi=37.5\;\GHz$, (b) $\rabi/2\pi=75.0\;\GHz$, and (c) experimental data (interaction strength is estimated as $\rabi/2\pi=150.1\;\GHz$) are plotted as a function of frequency $\omega/2\pi$ and external static magnetic flux density $\Bstat$. Dashed red and black lines show bare cavity and CR frequencies, respectively, without considering their interaction. At $\Bstat>0$, the CR and a circularly polarized probe THz wave are corotating and the vacuum Rabi splitting (anticrossing) is obtained. At $\Bstat<0$, the vacuum BS shift appears due to the counter-rotating coupling between the CR and the circularly polarized cavity field as highlighted by the gray shaded areas, even in the vacuum limit. The calculations in (a) and (b) were performed by the transfer matrix method by considering the sample structure depicted in Fig.~\ref{intro}(a). The (thickness, $\diecav$) of the air, mirror Si, defect Si layers are $(193.0\;\micron, 1+0.03\ii)$, $(50.0\;\micron, 3.42^2)$, and $(101.1\;\micron, 3.42^2)$, respectively. The thickness of the 2DEG is $d_{\text{2DEG}} = 2.31\;\micron$. Its relative dielectric function is $\diebg+\ii\sigma_{\text{AC}}(\omega)/(\omega\diez d_{\text{2DEG}})$, where $\diebg=3.6^2$ and the surface AC conductivity is $\sigma_{\text{AC}}(\omega)=\sigma_{\text{DC}}/[1-\ii(\omega\mp\wcyc)\tau]$ for surface DC conductivity $\sigma_{\text{DC}}=e\dens\mu$, mobility $\mu=9.0\times10^6\;\mathrm{cm^2/Vs}$, relaxation time $\tau=m\mu/e$, effective mass $m=0.067m_0$, where $m_0$ is the free electron mass in vacuum. Concerning the electron surface density, (c) $\dens=3.2\times10^{12}\;\mathrm{cm^{-2}}$ was estimated by a Hall measurement at 300\;mK, and those assumed in (a) and (b) are $\dens/16$ and $\dens/4$, respectively.\label{VBS}}
\end{figure*}
%%%%%%%%%%%%%%%%%%%%%%%%%%%%%%%%%%%%%%%%%%%%%%%%%%%%%%%%

Figures~\ref{VBS}(a) and (b) show numerically calculated THz transmission spectra for different values of the static magnetic flux density $\Bstat$. The interaction strengths are assumed to be (a) $\rabi/2\pi=37.5\;\GHz$ and (b) $\rabi/2\pi=75.0\;\GHz$ (see the details in the caption). The dashed red and black lines show the resonance frequencies of the cavity mode and CR, respectively, without considering their interaction. Figure~\ref{VBS}(c) shows experimental data, where we estimated $\rabi/2\pi = 150.1\;\GHz$. We used a circularly polarized THz wave as a probe beam, and it was sufficiently weak so that any optical nonlinear effects did not appear. At $\Bstat > 0$, the CR and the THz wave are corotating, and an anticrossing (vacuum Rabi splitting) appears. In contrast, at $\Bstat < 0$, the CR is counter-rotating against the THz wave, and a frequency shift is obtained, instead of the anticrossing, as highlighted with the gray shaded areas.

These anticrossing and frequency shift can be reproduced by the Hamiltonian $\oHH_{\text{Landau}}$ in Eq.~\eqref{eq:oH_Landau}. In its third term, the CR interacts with the $+$ and $-$ circularly polarized cavity modes in the corotating manner $\ii\hbar\rabi(\obd\oa_+-\oad_+\ob)$ and the counter-rotating manner $\ii\hbar\rabi(\obd\oad_--\oa_-\ob)$, respectively. This aspect is clearly reflected in the eigenfrequencies $\omega_{\pm}$ of the coupled modes (polariton modes) with the $\pm$ circular polarization determined by
\begin{subequations}
\begin{align}
\frac{\wA{}^2}{\omega_\pm{}^2}
& = 1 - \frac{2\rabit^2}{\omega_\pm{}^2} \mp \frac{2\rabit{}^2}{\omega_\pm(\omega_\pm\mp\wcyc)}\frac{\wcyc}{\omega_\pm} \label{eq:dispersion_A2_LM}\\
& = 1 - \frac{2\rabit^2}{\omega_\pm(\omega_\pm\mp\wcyc)}. \label{eq:dispersion_sucept}
\end{align}
\end{subequations}
Here, $\rabit \equiv \rabi\sqrt{\wA/\wcyc}$. These equations are obtained from the Fourier transform of the Heisenberg equations derived from $\oHH_{\text{Laudau}}$. The second term in Eq.~\eqref{eq:dispersion_sucept} corresponds to the optical susceptibility. Since it is proportional to $(\omega_+-\wcyc)^{-1}$, the $+$ circularly polarized photon resonantly interacts with the CR, and the anticrossing is obtained as seen in Fig.~\ref{VBS} for $\Bstat >0$. On the other hand, the $-$ circularly polarized mode does not show such an anticrossing since the optical susceptibility is proportional to $(\omega_-+\wcyc)^{-1}$. Since this frequency dependence purely comes from the counter-rotating coupling between CR and $-$ circularly polarized photons, the frequency shift observed for $\Bstat<0$ in Fig.~\ref{VBS} (gray shaded areas) can be interpreted as a BS shift. In contrast to the standard BS shift proportional to the intensity of the EM wave, the frequency shift in our system depends on the vacuum fluctuation $\braket{\oA_{\xi}(\pos)\oA_{\xi}(\pos)} = \hbar f(\pos)^2/[2\diez\diecav(\pos)\wA\area]= \hbar\rabit^2m/(Ne^2\wA)$ of the EM field, and it is interpreted as the vacuum counterpart of the BS shift.

In this way, the contributions of the corotating and counter-rotating couplings can be distinguished through the circular polarization in the CR system. Such a distinguishability cannot be obtained for the Lorentz-type susceptibility
\begin{equation}
\frac{\wA{}^2}{\omega^2}=1+\frac{4\rabit{}^2}{\wB{}^2-\omega^2},
\end{equation}
which is derived from the Hopfield Hamiltonian, Eq.~\eqref{eq:oH_Hopfield}, for $\AA=\rabi^2/\wB$ and $\rabit=\rabi\sqrt{\wA/\wB}$. In this case, the corotating contribution $(\wB-\omega)^{-1}$ and the counter-rotating one $(\wB+\omega)^{-1}$ cannot be clearly separated in the linear optical spectra.

On the other hand, the second term in Eq.~\eqref{eq:dispersion_A2_LM} comes from the $A^2$ term. In the case of $\Bstat = 0$, we get $\wcyc=\rabi=0$ while keeping $\rabit\neq0$. Then, the last term in Eq.~\eqref{eq:dispersion_A2_LM} disappears, and we get $\omega_{\pm}(\Bstat=0) = \sqrt{\wA{}^2+2\rabit^2}$, whose frequency shift $\Delta\omega \equiv \omega_{\pm}(\Bstat=0)-\wA$ from the bare cavity frequency $\wA$ is purely the contribution of the $A^2$ term. In the limit of $\Bstat\to-\infty$, this frequency shift is canceled by the vacuum BS shift as seen in Fig.~\ref{VBS}. In this way, the contribution of the $A^2$ term is also clearly distinguished from those of the corotating and counter-rotating couplings in the CR system. 

The requirement for observing the vacuum Rabi splitting is that $2\rabi$ should be larger than the line broadening. Owing to this relatively easy requirement, the ultrastrong vacuum Rabi splitting has been observed in a variety of systems~\cite{Forn-DiazetAl18arXiv,KockumetAl18arXiv}. In contrast, the maximum vacuum BS shift is $\Delta\omega=\sqrt{\wA{}^2+2\rabit^2}-\wA$, and it is approximately expressed as $\Delta\omega\approx\rabit^2/\wA$ in the case of $\rabit\ll\wA$. Therefore, in order to observe the vacuum BS shift, in addition to the ultrastrong $\rabi$ in the CR system and the circularly polarized THz wave, the high-$Q$ THz cavity and the high-mobility 2EDG are essential for realizing small enough line broadening as shown in Fig.~\ref{VBS}(c). These conditions were all satisfied in our experiments~\cite{LietAl18NP}.

%Figure~\ref{VBS} depicts how a vacuum Bloch-Siegert shift appears in a cavity system in the ultrastrong coupling regime without a strong light field. Figures\,\ref{VBS}(a), (b), and (c) show system eigenfrequencies with $g/2\pi=37.5$~GHz, $g/2\pi=75.0$~GHz, and $g/2\pi=150.1$~GHz, respectively, while controlling all other material and structural parameters to be the same and, most importantly, THz probe light is weak so that the average photon number is close to zero. Fig.~\ref{VBS}(a) and (b) are calculations from Eq.~\eqref{eq:oH_Landau}. Figures\,\ref{VBS}(c) is experimental data. The counter-rotating terms only affect the mode frequency in the negative magnetic field region, contributing to vacuum Bloch-Siegert shifts highlighted by the grey shaded areas marked om all panels. We see that the vacuum Bloch-Siegert shift can only be observed when $g$ is ultralarge, and the mode linewidths are sharp enough to provide high enough frequency resolution; both conditions are satisfied in our experiment shown by in Fig.~\ref{VBS}(c).

%%%%%%%%%%%%%%%%%%%%%%%%%%%%%%%%%%%%%%%%%%%%%%%%%%%%%%%%%%%%%
\section{Cooperative interaction in magnetic materials} \label{sec:ErFeO3}
\def\ee{\mathrm{e}}
\def\ii{\mathrm{i}}
\def\dd{\mathrm{d}}

\def\ddt#1{\frac{\partial #1}{\partial t}}

\def\Hc{\text{H.c.}}
\def\muB{\mu_{\text{B}}}
\def\kB{k_{\text{B}}}
\def\Hm{H^{\text{eff-mag}}}
\def\wFM{\omega_{\text{FM}}}
\def\wCFT{\omega_{\text{CFT}}}

\def\HH{\mathcal{H}}
\def\HHmag{\HH_{\text{mag}}}
\def\HHmagSimple{\HH_{\text{mag}}^{\text{simple}}}
\def\HHint{\HH_{\text{int}}}

\def\oHH{\hat{\mathcal{H}}}
\def\oHHLZ{\oHH_{\text{LZ}}}
\def\oHHQZ{\oHH_{\text{QZ}}}
\def\oHHFeEr{\oHH_{\text{Fe--Er}}}
\def\oHHmag{\oHH_{\text{mag}}}
\def\oHHFM{\oHH_{\text{FM}}}
\def\oHHEPR{\oHH_{\text{EPR}}}
\def\oHHint{\oHH_{\text{int}}}

\def\osigma{\hat{\sigma}}
\def\oH{\hat{H}}
\def\oa{\hat{a}}
\def\oad{\hat{a}^{\dagger}}
\def\ob{\hat{b}}
\def\obd{\hat{b}^{\dagger}}
\def\omu{\hat{\mu}}
\def\oS{\hat{S}}
\def\oT{\hat{T}}
\def\oY{\hat{Y}}
\def\ovmu{\hat{\bm{\mu}}}
\def\ovR{\hat{\bm{R}}}
\def\ovS{\hat{\bm{S}}}

\def\vunit{\bm{e}}
\def\vH{\bm{H}}
\def\vHext{\bm{H}_{\text{ext}}}
\def\vHeff{\bm{H}_{\text{eff}}}
\def\vD{\bm{D}}
\def\vSa{\bar{\bm{S}}}
\def\Sa{\bar{S}}
\def\vS{\bm{S}}
\def\Hext{H_{\text{ext}}}
\def\Heffx{H_{\text{eff},x}}
\def\Heffy{H_{\text{eff},y}}
\def\Heffz{H_{\text{eff},z}}

\def\HHFe{\mathcal{H}_{\text{Fe}}}
\def\HHEr{\mathcal{H}_{\text{Er}}}
\def\HHFeEr{\mathcal{H}_{\text{Fe--Er}}}

\def\oHHFe{\hat{\mathcal{H}}_{\text{Fe}}}
\def\oHHEr{\hat{\mathcal{H}}_{\text{Er}}}
\def\oHHFeEr{\hat{\mathcal{H}}_{\text{Fe--Er}}}

\def\JFe{J_{\text{Fe}}}
\def\DFe{D_{\text{Fe}}}
\def\Ax{A_x}
\def\Az{A_z}
\def\Axz{A_{xz}}
\def\NUC{N_{\text{UC}}}
\def\wFM{\omega_{\text{FM}}}
\def\wAFM{\omega_{\text{AFM}}}
\def\wEPR{\omega_{\text{EPR}}}
\def\shift{\lambda}

\def\gf{\mathfrak{g}}
\def\Tc{T_{\text{c}}}
\def\vS{\bm{S}}
\def\vR{\bm{R}}

\def\sumnn{\sum_{\mathrm{n.n.}}}
\def\etaspin{\eta_{\text{spin}}}
\def\meV{\mathrm{meV}}
\def\THz{\mathrm{THz}}

In this section, we will derive the Dicke Hamiltonian, Eq.~\eqref{eq:oH_ErFeO3}, from the following spin model for $\mathrm{ErFeO_3}$:
\begin{equation} \label{eq:spin_model} % !!!!!!!!!!!!!!!!!!!!!!!!!!!!!!!!!!!!!
\oHH = \oHHFe + \oHHEr + \oHHFeEr.
\end{equation}
Following the discussion by G.~F.~Herrmann in 1963~\cite{Herrmann63JPCS}, we express the Hamiltonian of the $\Feion$ spins by a two-sublattice model as
\begin{widetext}
\begin{align} \label{eq:VFe} % !!!!!!!!!!!!!!!!!!!!!!!!!!!!!!!!!!!!!!!!!!
\oHHFe
& = \JFe \sumnn \ovS_i^A \cdot \ovS_{i'}^B
- \DFe \sumnn \left( \oS_{i,z}^A \oS_{i',x}^B - \oS_{i',z}^B \oS_{i,x}^A \right)
- \sum_{i=1}^N \left( \Ax \oS_{i,x}^A{}^2 + \Az \oS_{i,z}^A{}^2 + \Axz \oS_{i,x}^A \oS_{i,z}^A \right)
\nonumber \\ & \quad
- \sum_{i=1}^N \left( \Ax \oS_{i,x}^B{}^2 + \Az \oS_{i,z}^B{}^2 - \Axz \oS_{i,x}^B \oS_{i,z}^B \right).
\end{align}
\end{widetext}
Here, $\ovS_i^{A/B}$ is the operator of the $i$-th $\Feion$ spin vector with magnitude $S=5/2$ in its $A/B$ sublattice. $\sumnn$ means a summation over all the nearest neighbor couplings. $N$ is the number of $\Feion$ spins in each sublattice, i.e., there are in total $2N$ spins representing the $\Feion$ subsystem. $\JFe$ and $\DFe$ are the isotropic and antisymmetric exchange constants, respectively. $\Ax$, $\Az$, and $\Axz$ are the magnetic anisotropy energies.
On the other hand, we assume that the Hamiltonian of the $\Erion$ spins is simply expressed under an external static magnetic flux density $\vBstat$ as
\begin{equation}
\oHHEr = - \sum_{i=1}^N \ovmu_i\cdot \vBstat.
\end{equation}
Here, $\ovmu_i$ is the operator of the $i$-th $\Erion$ magnetic moment vector expressed by the Pauli operators $\{\osigma_{i,\xi}\}$ as $\ovmu_i = - \muB (\gf_x \osigma_{i,x}, \gf_y \osigma_{i,y}, \gf_z \osigma_{i,z})^{\mathrm{t}}$, where $\gf_\xi$ is the anisotropic $g$-factor. In the following, the $\Erion$ spin vector is expressed as $\ovR_i = (\osigma_{i,x}, \osigma_{i,y}, \osigma_{i,z})^{\text{t}}$. We do not consider the $\Erion$--$\Erion$ exchange interactions. Finally, the $\Feion$--$\Erion$ exchange interactions are expressed as
\begin{widetext}
\begin{equation} \label{eq:VFeEr} % !!!!!!!!!!!!!!!!!!!!!!!!!!!!!!!!!!!!!!!!!!
\oHHFeEr = \sum_{i=1}^N \left[
  J_A \ovR_i \cdot \ovS_i^A + J_B \ovR_i \cdot \ovS_i^B
+ \vD_A \cdot (\ovR_i \times \ovS_i^A) + \vD_B \cdot (\ovR_i \times \ovS_i^B)
\right].
\end{equation}
\end{widetext}
Here, $J_{A/B}$ and $\vD_{A/B}$ are the symmetric and antisymmetric exchange constants, respectively. In this way, we assume that the $\Feion$ spins in the $i$-th unit cell of $\mathrm{ErFeO_3}$ are represented by $\ovS^{A/B}_i$ (two-sublattice model), each of which is in fact a sum of two real $\Feion$ spins in the four-sublattice model. On the other hand, in the same manner as Ref.~\citenum{LietAl18Science}, the four $\Erion$ spins in the $i$-th unit cell are represented simply by one Pauli vector $\ovR_i$.

\begin{figure}%[b]
\begin{center}
\begin{tabular}{c}
\includegraphics[scale=.6]{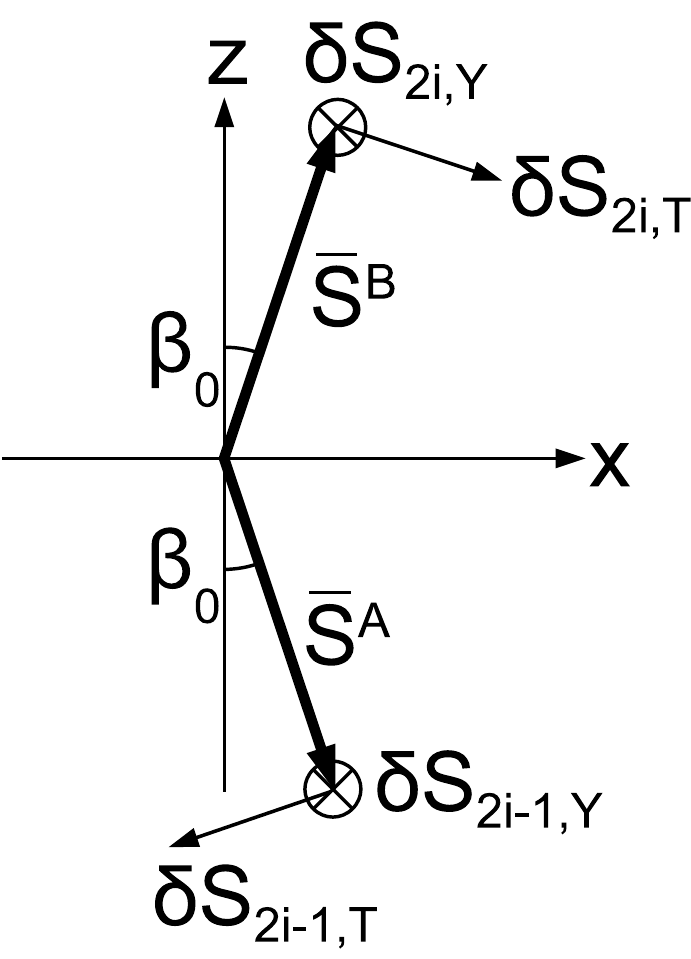}
\end{tabular}
\end{center}
\caption{Equilibrium spins $\vSa^{A/B}$ of $\Feion$ are antiferromagnetically ordered along the $z$ axis, while they are canted toward the $x$ direction by angle $\beta_0$. The magnons are described as a wave of modulations $\{\delta\vS_{\ell,T/Y}\}$ from them, where $\delta\vS_{2i,T/Y}$ and $\delta\vS_{2i-1,T/Y}$ are the modulations of the two spins in the $i$-th unit cell.\label{fig:1}}
\end{figure} 

In the ground state of the $\Feion$ subsystem, their spins are ordered antiferromagnetically along the $z$ axis as depicted in Fig.~\ref{fig:1}, but slightly canted toward the $x$ direction by angle $\beta_0 = (-1/2)\arctan[(\Axz+\DFe)/(\JFe-\Ax+\Az)]$~\cite{Herrmann63JPCS}. The equilibrium spins are expressed as $\vSa^A_i=S(\sin\beta_0, 0, -\cos\beta_0)^{\mathrm{t}}$ and $\vSa^B_i=S(\sin\beta_0, 0, \cos\beta_0)^{\mathrm{t}}$. We define the modulations $\{\delta\oS_{\ell,T},\delta\oS_{\ell,Y}\}$ from them as depicted in Fig.~\ref{fig:1}, and these operators satisfy $[\delta\oS_{\ell,T}, \delta\oS_{\ell',Y}] = \ii\delta_{\ell,\ell'}$. The spin modulations are then expressed as
\begin{subequations}
\begin{align}
\delta\ovS^A_i &= \ovS^A_i - \vSa^A_i
= \begin{pmatrix} -\delta\oS_{2i-1,T}\cos\beta_0 \\ \delta\oS_{2i-1,Y} \\ -\delta\oS_{2i-1,T}\sin\beta_0 \end{pmatrix}, \\
\delta\ovS^B_i &= \ovS^B_i - \vSa^B_i
= \begin{pmatrix} \delta\oS_{2i,T}\cos\beta_0 \\ \delta\oS_{2i,Y} \\ -\delta\oS_{2i,T}\sin\beta_0 \end{pmatrix}.
\end{align}
\end{subequations}
Extending Herrmann's calculation~\cite{Herrmann63JPCS} into a magnon model propagating in the $z$ direction (under averaging in the $x$--$y$ plane)~\cite{TsangetAl78JAP}, we can derive the equations of motion for these modulations as
\begin{subequations} \label{eq:motion_magnon} % !!!!!!!!!!!!!!!!!!!!!!!!!!!!
\begin{align}
\frac{1}{\gamma}\ddt{}\delta\oS_{\ell,T}
& = - a \delta\oS_{\ell,Y}
    + \frac{b}{2}\left( \delta\oS_{\ell-1,Y} + \delta\oS_{\ell+1,Y} \right), \\
\frac{1}{\gamma}\ddt{}\delta\oS_{\ell,Y}
& = - c \delta\oS_{\ell,T}
    - \frac{d}{2}\left( \delta\oS_{\ell-1,T} + \delta\oS_{\ell+1,T} \right).
\end{align}
\end{subequations}
Here, $\gamma = \gf \muB /\hbar$ is the gyromagnetic ratio for the free electron $g$-factor $\gf$ and the Bohr magneton $\muB$. The coefficients are expressed as~\cite{Herrmann63JPCS}
\begin{widetext}
\begin{subequations} \label{eq:abcd} % !!!!!!!!!!!!!!!!!!!!!!!!!!!!!!!!!!!!!!!
\begin{align}
a & = [S/(g\muB)] [ - \Az - \Ax - (z\JFe + \Az - \Ax)\cos(2\beta_0) + (\Axz+z\DFe)\sin(2\beta_0)], \\
b & = [S/(g\muB)] (z\JFe), \\
c & = [S/(g\muB)] [(z\JFe+2\Az-2\Ax)\cos(2\beta_0) + z\DFe\sin(2\beta_0)], \\
d & = [S/(g\muB)] [-z\JFe\cos(2\beta_0) - (2\Axz+z\DFe)\sin(2\beta_0)],
\end{align}
\end{subequations}
\end{widetext}
where $z = 6$ is the number of neighboring $\Feion$ sites for each $\Feion$ spin. Then, the Hamiltonian of the $\Feion$ spins is approximated (bosonized) as
\begin{widetext}
\begin{equation} \label{eq:H_magnon} % !!!!!!!!!!!!!!!!!!!!!!!!!!!!!!!!!!!!!!!
\oHHFe
\approx \hbar\gamma \sum_{\ell=1}^{2N_z} \left(
  - \frac{a}{2} \delta\oS_{\ell,Y}{}^2
  + \frac{c}{2} \delta\oS_{\ell,T}{}^2
  + \frac{b}{2} \delta\oS_{\ell,Y} \delta\oS_{\ell+1,Y}
  + \frac{d}{2} \delta\oS_{\ell,T} \delta\oS_{\ell+1,T}
  \right) + \const
\end{equation}
\end{widetext}
Here, $N_z$ and $2N_z$ are the number of unit cells and of $\Feion$ spins, respectively, in the $z$ direction. In terms of the annihilation operator $\oa_K$ of a magnon with a dimensionless wavenumber $K$, satisfying $[\oa_K, \oad_{K'}] = \delta_{K,K'}$, the modulation operators are expressed as
\begin{subequations}
\begin{align}
\delta\oS_{\ell,T}
& = \sqrt{\frac{1}{2N_z}} \sum_{K=-\pi}^\pi \ee^{\ii K \ell}
    \left(\frac{b\cos(K)-a}{d\cos(K)+c}\right)^{1/4}
    \frac{\oad_{-K} - \oa_{K}}{\ii\sqrt{2}}, \\
\delta\oS_{\ell,Y}
& = \sqrt{\frac{1}{2N_z}} \sum_{K=-\pi}^\pi \ee^{\ii K \ell}
    \left(\frac{d\cos(K)+c}{b\cos(K)-a}\right)^{1/4}
    \frac{\oad_{-K} + \oa_{K}}{\sqrt{2}}.
\end{align}
\end{subequations}
The Hamiltonian in Eq.~\eqref{eq:H_magnon} is rewritten as
\begin{align} \label{eq:oHHFe2} % !!!!!!!!!!!!!!!!!!!!!!!!!!!!!!!!!!!!!!!!!!!!
\oHHFe
& \approx \sum_{K=-\pi}^\pi \hbar\omega_K \left(
    \oad_{K}\oa_{K} + \frac{1}{2}
  \right) + \const,
\end{align}
where the eigenfrequency is expressed as
\begin{equation}
\omega_K = \gamma\sqrt{[b\cos(K)-a][d\cos(K)+c]}.
\end{equation}
Here, $K=0$ and $K=\pi$ correspond to the qFM and quasi-antiferromagnetic (qAFM) modes~\cite{Herrmann63JPCS}, respectively.

%%%%%%%%%%%%%%%%%%%%%%%%% figure 3 %%%%%%%%%%%%%%%%%%%%%
\begin{figure*}
\begin{center}
\begin{tabular}{c}
\includegraphics[width=0.85\linewidth]{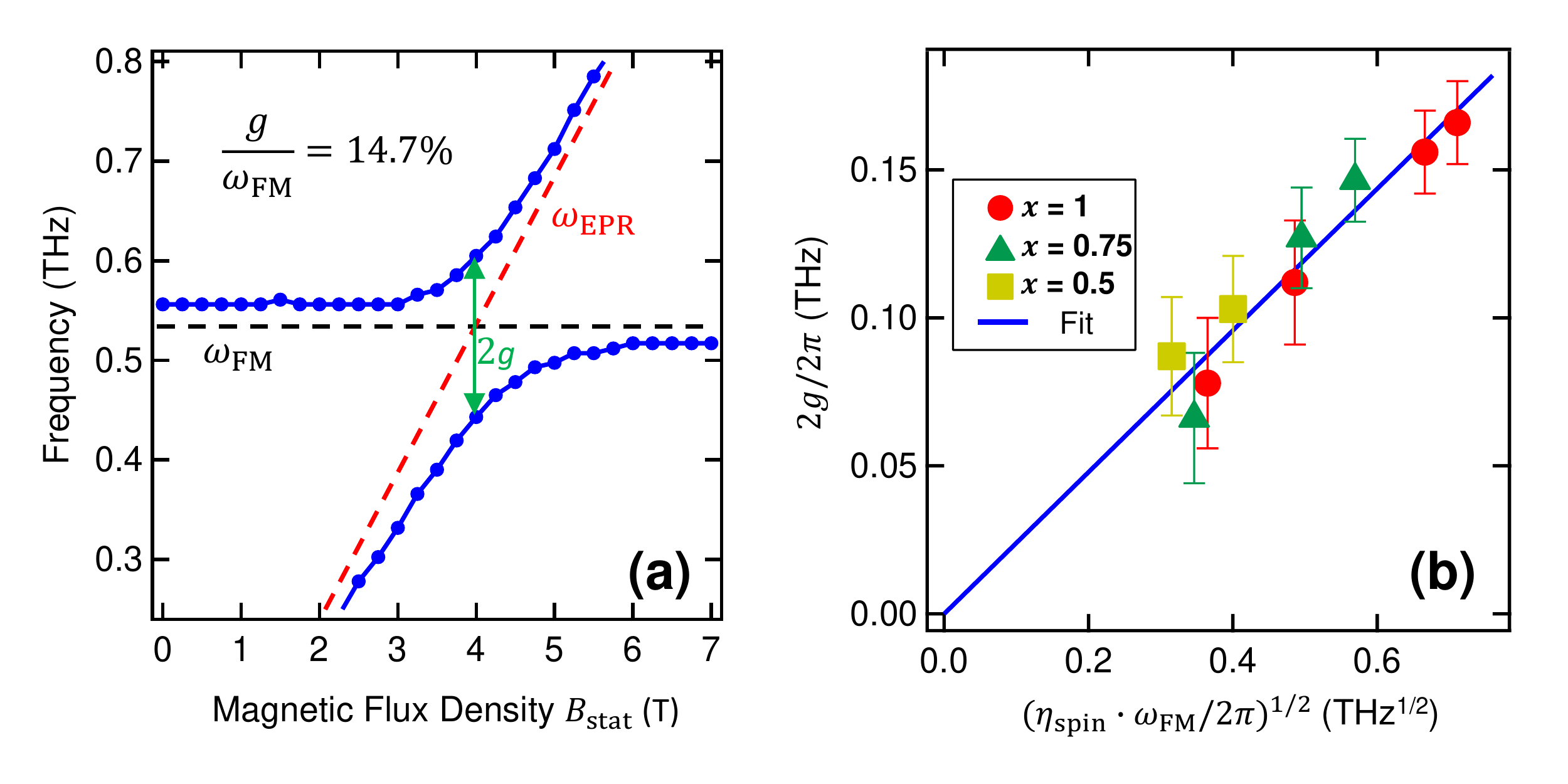}
\end{tabular}
\end{center}
\caption{Experimentally observed absorption peak frequencies of $\mathrm{ErFeO_3}$ at 10~K under an external static magnetic field along the $c$ axis. The probe THz wave was propagating along the $c$ axis. Dashed red and black lines show uncoupled EPR and magnon frequencies, respectively. (b) Cooperative scaling behavior of $\rabi$ as a function of $(\eta_{\text{spin}}\omega_{\text{FM}})^{1/2}$, where $\eta_{\text{spin}}$ is the net fraction of EPR-contributing $\mathrm{Er}^{3+}$ spins.\label{ErFO}}
\end{figure*}
%%%%%%%%%%%%%%%%%%%%%%%%%%%%%%%%%%%%%%%%%%%%%%%%%%%%%%%%

Dividing the $\Feion$ spin operators into the equilibrium value $\vSa^{A/B}$ and the modulation operators $\{\delta\ovS^{A/B}_i\}$, the rest of the Hamiltonian is rewritten as
\begin{equation}
\oHHEr + \oHHFeEr = \oHHEPR + \oHHint,
\end{equation}
where
\begin{widetext}
\begin{align}
\oHHEPR
& = \sum_{i=1}^{N_z} \left[ - \ovmu_i \cdot \vBstat
  + J_A \ovR_i\cdot\vSa^{A} + J_B \ovR_i\cdot\vSa^{B}
  + \vD_A \cdot(\ovR_i\times\vSa^{A}) + \vD_B \cdot(\ovR_i\times\vSa^{B}) 
\right], \label{eq:oHHCFT} \\ % !!!!!!!!!!!!!!!!!!!!!!!!!!!!!!!!!!!!!!!!!!!!!!!
\oHHint
& = \sum_{i=1}^{N_z} \left[
    J_A \ovR_i\cdot\delta\ovS_{i}^A + J_B \ovR_i\cdot\delta\ovS_{i}^B
  + \vD_A \cdot(\ovR_i\times\delta\ovS_{i}^A)
  + \vD_B \cdot(\ovR_i\times\delta\ovS_{i}^B) 
\right]. \label{eq:oHHint} % !!!!!!!!!!!!!!!!!!!!!!!!!!!!!!!!!!!!!!!!!!!!!!!!!!
\end{align}
\end{widetext}
We also averaged the $\Erion$ spins in the $x$--$y$ plane. In the experiment of Ref.~\citenum{LietAl18Science}, the main contribution to the EPR of $\Erion$ ions was the the external static magnetic field along the $c$ ($z$) axis, and the exchange contribution from the equilibrium $\Feion$ spins $\vSa^{A/B}$ was a minor one. Then, the EPR Hamiltonian is approximated as
\begin{equation} \label{eq:oHHEPR} % !!!!!!!!!!!!!!!!!!!!!!!!!!!!!!!!!!!!!!!!!
\oHHEPR \approx \hbar\wEPR \sum_{i=1}^{N_z}\frac{\osigma_{i,z}}{2},
\end{equation}
where $\wEPR = 2g_z \muB \Bstat/\hbar$. A more detailed analysis was performed in the Supplementary Materials of Ref.~\citenum{LietAl18Science}. On the other hand, the EPR--magnon interaction is dominated by the symmetric exchange interactions, since $J_{A/B} \gg |\vD_{A/B}|$ is usually satisfied. Due to a symmetry analysis, we can find $J_A=J_B=J$~\cite{LietAl18Science}. Then, when we assume the long-wavelength limit and focus only on the qFM magnon mode with frequency $\wFM = \omega_{K=0}$, which shows the anticrossing with EPR as seen in Fig.~\ref{ErFO}(a), the interaction Hamiltonian is approximated as
\begin{widetext}
\begin{equation}
\oHHint \approx \sum_{i=1}^{N_z} J \ovR_i\cdot(\delta\ovS_{i}^A + \delta\ovS_{i}^B)
\approx J \sqrt{\frac{1}{N_z}} \sum_{i=1}^{N_z} \left[
    \left(\frac{d+c}{b-a}\right)^{1/4}
    \osigma_{i,y} \left( \oad_0 + \oa_0 \right)
  + \ii\left(\frac{b-a}{d+c}\right)^{1/4} \sin(\beta)
    \osigma_{i,z} \left( \oad_0 - \oa_0 \right)
  \right].
\end{equation}
\end{widetext}
Since the last term gives a nonlinear optical response for the approximated EPR Hamiltonian in Eq.~\eqref{eq:oHHEPR}, the interaction Hamiltonian is approximated in the linear responses as
\begin{equation} \label{eq:oHHint2} % !!!!!!!!!!!!!!!!!!!!!!!!!!!!!!!!!!!!!!!!!
\oHHint
\approx \frac{\hbar\rabi}{\sqrt{\etaspin N_z}} \sum_{i=1}^{\etaspin N_z}
    \osigma_{i,y} \left( \oad_0 + \oa_0 \right).
\end{equation}
In experiments, we can dilute the density of $\Erion$ ions on demand by substitutional doping with nonmagnetic $\mathrm{Y^{3+}}$ ions. Considering the dilution of $\Erion$ spins by factor $x$ due to the replacement with $\mathrm{Y}^{3+}$ ions and also the thermal distribution of electron spins in $\Erion$ ions, the net fraction $\etaspin \equiv x \tanh(-\hbar\wEPR/\kB T)$ of the EPR-contributing $\Erion$ density was introduced in Eq.~\eqref{eq:oHHint2}~\cite{LietAl18Science}. The interaction strength can be expressed as
\begin{equation} \label{eq:rabi_ErFeO3} % !!!!!!!!!!!!!!!!!!!!!!!!!!!!!!!!!!!!!!!
\rabi
\equiv \frac{J}{\hbar} \sqrt{\etaspin} \left(\frac{d+c}{b-a}\right)^{1/4}
\approx \frac{J}{\hbar} \sqrt{\frac{\etaspin\wFM}{2zS\JFe/\hbar}}.
\end{equation}
In this way, the interaction strength $\rabi$ is proportional to the square root of $\etaspin$. From Eqs.~\eqref{eq:oHHFe2}, \eqref{eq:oHHEPR}, and \eqref{eq:oHHint2}, we get the Dicke Hamiltonian, i.e., Eq.~\eqref{eq:oH_ErFeO3}. Note that no $A^2$ term appears in this Dicke Hamiltonian (even without the renormalization mentioned in \cite{FN1}, 
since the magnon--EPR interaction is derived from the $\Erion$--$\Feion$ exchange interactions in the spin model of Eq.~\eqref{eq:spin_model}, not from the kinetic energy in the minimal-coupling Hamiltonian.

Figure \ref{ErFO}(a) shows experimentally observed absorption peak frequencies of $\mathrm{ErFeO_3}$ at 10\;K as a function of static magnetic flux density $\Bstat$ along the $c$ ($z$) axis. The dashed red and black lines show uncoupled EPR and magnon frequencies, respectively. The interaction strength $\rabi$ can be directly read from the vacuum Rabi splitting $2\rabi$ in the graph, and we found $\rabi/\wFM = 14.7\;\%$, reaching the ultrastrong coupling regime. Moreover, by measuring multiple samples of $\mathrm{Er}_x\mathrm{Y}_{1-x}\mathrm{FeO_3}$ at multiple temperatures, we found that $\rabi$ showed a proportionality with $(\etaspin\wFM)^{1/2}$ as shown in Fig.~\ref{ErFO}(b) and as theoretically derived in Eq.~\eqref{eq:rabi_ErFeO3}. This proportionality is experimental evidence of the Dicke cooperativity between the ensemble of $\Erion$ ions and the magnon mode in the $\Feion$ lattice. While the origin of their interaction is short-range (nearest-neighboring) exchange interactions as modeled in Eq.~\eqref{eq:VFeEr}, the EPR of the $\Erion$ ensemble cooperatively interacts with the $\Feion$ magnon mode, which propagates through the $\Feion$--$\Feion$ exchange interactions.

While $\wFM$ were slightly modified depending on the sample and temperature, we assume $\JFe = 4.96\;\meV$ as reported in Ref.~\citenum{KoshizukaHayashi88JPSJ}. From the proportionality $2\rabi/(2\pi) \approx (\etaspin\wFM/2\pi)\times0.238\;\THz^{1/2}$ obtained in Fig.~\ref{ErFO}(b), we can estimate the $\Feion$--$\Erion$ symmetric exchange interaction strength as $J = 2.95\;\meV$.

%$\DFe = -0.107\;\meV$,\cite{Koshizuka1988} $\Ax=0.0076\;\meV$, $\Az = 0.015\;\meV$, $\Axz = 0$, we get $\wFM/2\pi = 0.567\;\THz$ but modified by ...

\section{Summary} \label{sec:summary}
The counter-rotating light--matter coupling and the $A^2$ (quadratic) term are essential for exploring the virtual photons and the SRPT in the USC regime. In the CR system, apart from the contribution of the corotating coupling (vacuum Rabi splitting), those of the counter-rotating coupling and $A^2$ term can be clearly distinguished, respectively, as the vacuum BS shift and the cavity frequency shift at zero static magnetic field. In a bulk $\mathrm{ErFeO_3}$ crystal, the coupled system of the $\Erion$ spins and $\Feion$ magnons can be described by the Dicke Hamiltonian. If we find magnon analogues of the SRPT in such magnetic materials, it would give us a hint of realizing the original photon SRPT. The possibility of SRPT still remains in the light--spin coupling systems experimentally, while the thermal SRPT has been discussed mostly in ideal theoretical models. The systems without the time reversal symmetry and a variety of magnetic materials would advance the exploration of virtual photons and the SRPT in the USC regime.

%%%%%%%%%%%%%%%%%%%%%%%%%%%%%%%%%%%%%%%%%%%%%%%%%%%%%%%%%%%%%
%\acknowledgments     %>>>> equivalent to \section*{ACKNOWLEDGMENTS}       
\section*{ACKNOWLEDGMENTS} 

This research was supported by the National Science Foundation (Cooperative Agreement DMR-1720595), the U.S. Army Research Office (grant W911NF-17-1-0259), the JST PRESTO program (grant JPMJPR1767), KAKENHI (grant 26287087), and the ImPACT Program of the Council for Science, Technology and Innovation (Cabinet Office, government of Japan). We thank Katsumasa Yoshioka for his help with all measurements done with circularly polarized THz radiation. Saeed Fallahi, Geoff Gardner, and Michael Manfra provided the 2DEG sample for the Landau polariton work in Section \ref{sec:Landau}. Ning Yuan, Maolin Xiang, Kai Xu, and Shixun Cao provided the $\mathrm{Er}_x\mathrm{Y}_{1-x}\mathrm{FeO}_3$ crystals for the work described in Section \ref{sec:ErFeO3}.

%%%%%%%%%%%%%%%%%%%%%%%%%%%%%%%%%%%%%%%%%%%%%%%%%%%%%%%%%%%%%
%%%%% References %%%%%

%\bibliography{jun}   %>>>> bibliography data in report.bib

\begin{thebibliography}{35}
\bibitem{FN1} We can eliminate the $A^2$ term by a Bogoliubov transformation as $\oHH_{\text{Dicke}}=\hbar\wAt\oa^{\prime\dagger}\oa'+\hbar\wB\sum_{i=1}^N(\osigma_{i,z}/2) + (\ii\hbar\rabit/\sqrt{N})(\oa^{\prime\dagger}+\oa')\sum_{i=1}^N(\osigmad_i-\osigma_i) + \const$, where $\wAt \equiv \sqrt{\wA{}^2+4\AA\wA}$, $\rabit\equiv \rabi\sqrt{\wA/\wAt}$, and the photon operator is renormalized as $\oa'\equiv [(\wAt+\wA)\oa+(\wAt-\wA)\oad]/(2\sqrt{\wAt\wA})$. However, if $\AA\ge\rabi^2/\wB$ is satisfied, we always get $\rabit^2 < \wAt\wB/4$, and thus, the SRPT does not occur.
\bibitem{FN2} We can find that, in the case of $\rabi^2>\wA\wB/4$ and $\AA=0$, an eigenfrequency of the Hopfield Hamiltonian in Eq.~\eqref{eq:oH_Hopfield} becomes a pure imaginary value, which indicates the instability of the normal ground state. This is because the quartic part of the potential of the atomic collective excitation is neglected in $\oHH_{\text{Hopfield}}$, although it is essential for describing the double-well potential of the second-order phase transition in the Landau theory. When we consider the higher-order terms, the SRPT can be reproduced in the Hamiltonian after the Holstein-Primakoff transformation~\cite{EmaryBrandes03PRL,EmaryBrandes03PRE}.
\end{thebibliography}
%\bibliographystyle{spiebib}   %>>>> makes bibtex use spiebib.bst

\end{document}